\documentclass[11pt,twoside]{article}
\usepackage{maxim}

\def\all{\forall\,}
\def\wdt{\widetilde}
\def\cp{\operatorname{CP}}
\def\dd{{\sf D}}
\newcommand\tran[1]{#1^{\sf T}}
\newcommand\dnorm[1]{\norm{#1}_\lozenge}

\begin{document}
\pagestyle{myheadings}
\markboth{Maxim Raginsky}{Radon-Nikodym Derivatives of Quantum Operations}
\begin{titlepage}
\title{\bf Radon-Nikodym Derivatives of Quantum Operations}
\author{Maxim Raginsky\thanks{Electronic Mail:
    \texttt{maxim@ece.northwestern.edu}
}
  \\[1ex]
  {\small Center for Photonic Communication and Computing}\\
  {\small Department of Electrical and Computer Engineering}\\
  {\small Northwestern University, Evanston, IL 60208-3118, USA}}
\date{}

\maketitle

\begin{abstract}
\noindent{Given a completely positive (CP) map $T$, there is a theorem of the
  Radon-Nikodym type [W.B. Arveson, Acta Math. {\bf 123}, 141
  (1969); V.P. Belavkin and P. Staszewski, Rep. Math. Phys. {\bf 24}, 49
  (1986)] that completely characterizes all CP maps $S$
  such that $T-S$ is also a CP map. This theorem is reviewed, and
  several alternative formulations are given along the way. We then use
  the Radon-Nikodym formalism to study the structure of order intervals of
  quantum operations, as well as a certain one-to-one
  correspondence between CP maps and positive operators, already
  fruitfully exploited in many quantum information-theoretic
  treatments. We also comment on how the Radon-Nikodym theorem can be
  used to derive norm estimates for differences of CP maps
  in general, and of quantum operations in particular.}
\end{abstract}

\vfill

\small{

\noindent{\bf PACS Classification (2003):} 02.30.Tb, 03.67.-a

}

\setcounter{page}{0}
\thispagestyle{empty}
\end{titlepage}


\section{Introduction}
\label{sec:intro}

In the mathematical framework of quantum information theory
\cite{key}, all admissible devices are modelled by the so-called
quantum operations \cite{dav,kra} --- that is, completely positive
linear contractions on the algebra of observables of the physical system
under consideration.  Thus it is of paramount importance to have at
one's disposal a good analysis toolkit for completely positive (CP) maps.

There are many useful structure theorems for CP maps. The two best known
  ones, due to Stinespring \cite{sti} and Kraus
  \cite{kra}, are {\em de rigueur} in virtually all quantum
  information-theoretic treatments.  These theorems are significant
  because each of them states that a given map is CP if and only if it is
  expressible in a certain canonical form.  However, in many
  applications we need to consider whole families of CP maps. This
  necessitates the introduction of comparison tools for CP maps, e.g.,
  when the family of CP maps in question admits some sort of (partial) order.

Mathematically, the set of all CP maps between two algebras of
observables is a cone that can be partially ordered in the
following natural way. If $S$ and $T$ are two CP maps, we write $S \le
T$ if $T - S$ is CP as well.  This partial order comes up in, e.g.,
the problem of distinguishing between two known CP maps with given
{\it a priori} probabilities under the constraint that the average
probability of error is minimized \cite{dpp}.  A typical way of
dealing with partially ordered cones is to exhibit a correspondence
between the cone's order and a partial order of some ``simpler''
objects.  This is accomplished by means of theorems of the
Radon-Nikodym type, as in the case of, e.g., partial ordering of
positive measures or positive linear functionals.  There are a number of
Radon-Nikodym theorems for CP maps (see, e.g., the work of Arveson
\cite{arv},  Belavkin and Staszewski \cite{bs}, Davies \cite{dav},
Holevo \cite{hol}, Ozawa \cite{oza}, and Parthasarathy \cite{par})
that differ widely in scope and in generality. Thus, the
results of Davies, Ozawa, and Holevo have to do with Radon-Nikodym
derivatives of CP instruments \cite{oza} with respect to scalar
measures.  On the other hand, ideas common to the Arveson and
Belavkin-Staszewski theorems, with further developments by Parthasarathy,
are directly applicable to the partial ordering of CP maps described
above, and will therefore be the focus of the present article. More
specifically, we will demonstrate that certain problems encountered in
quantum information-theoretic settings that involve characterization
and comparison of CP maps, are best understood in this Radon-Nikodym framework.

The paper is organized as follows.  We
summarize the salient facts on CP maps and quantum
operations in Section \ref{sec:prelims}. In Section \ref{sec:rnt} we
review the Arveson-Belavkin-Staszewski formulation of the
Radon-Nikodym theorem for CP maps and state several alternative, but
equivalent, versions. The Radon-Nikodym machinery is then applied to
the following problems: partial ordering of quantum operations
(Section \ref{sec:poqo}), characterization of quantum operations by
means of positive operators (Section \ref{sec:cqopo}), and estimating norms of
differences of CP maps (Section \ref{sec:norms}). Finally some concluding
remarks are made in Section \ref{sec:rems}.

\section{Preliminaries}
\label{sec:prelims}

\subsection{Completely positive maps}
\label{ssec:cpmaps}

\paragraph{Definitions ---} Let $\sA$ and $\sB$ be C*-algebras; denote
by $\sA^+$ the cone of positive elements of $\sA$.  A linear map
$\map{T}{\sA}{\sB}$ is called {\em positive} if $T(\sA^+) \subseteq \sB^+$. Given some $n
\in \N$, let $\sM_n$ be the algebra of $n \times n$ complex
matrices. The map $T$ is called {\em $n$-positive} if the induced map
$\map{T \tp \id_n}{\sA \tp \sM_n}{\sB \tp \sM_n}$ is positive, and
{\em completely positive} if it is $n$-positive for all $n \in \N$.

One typically considers maps $\map{T}{\sA}{\sB(\sH)}$, where $\sA$ is
a C*-algebra with identity, and $\sB(\sH)$ is the algebra of bounded
operators on a complex separable Hilbert space $\sH$. Then it can
be shown \cite{sti} that $T$ is CP if and only if, for each $n \in \N$,
\begin{equation}
\sum^n_{i,j=1} \braket{\eta_i}{T(A^*_i A_j)\eta_j} \ge 0 \qquad \all
\eta_i \in \sH, A_i \in \sA; i = 1,\ldots, n.
\label{eq:cpcond}
\end{equation}

\paragraph{Theorems of Stinespring and Kraus ---} A fundamental
theorem of Stinespring \cite{sti} states that, for any normal (i.e.,
ultraweakly continuous) CP map $\map{T}{\sA}{\sB(\sH)}$, there exist a
Hilbert space $\sK$, a $*$-homomorphism $\map{\pi}{\sA}{\sB(\sK)}$,
and a bounded operator $\map{V}{\sH}{\sK}$, such that
\begin{equation}
T(A) = V^*\pi(A)V \qquad \all A \in \sA.
\label{eq:stiform}
\end{equation}
We will refer to any such triple $(\sK,V,\pi)$ [or, through a slight
  abuse of language, to the form (\ref{eq:stiform}) of $T$] as a {\em
  Stinespring dilation of $T$}. Given $T$, one can construct its Stinespring
  dilation in such a way that $\sK = \overline{\pi(\sA)V\sH}$, i.e.,
  the set $\setcond{\pi(A)V\psi}{A \in \sA,\psi \in \sH}$ is total in
  $\sK$. With this additional property, the Stinespring dilation is
  unique up to unitary equivalence \cite{el}, and is called the {\em
  minimal Stinespring dilation}.

For the special case of a CP map $\map{T}{\sB(\sH_1)}{\sB(\sH_2)}$, we
can always find a Hilbert space $\sE$ and a bounded operator
$\map{V}{\sH_2}{\sH_1 \tp \sE}$, such that 
\begin{equation}
T(A) = V^*(A \tp \idty_\sE)V \qquad \all A \in \sA.
\label{eq:stican}
\end{equation}
This follows from the fact that any normal $*$-representation of the
C*-algebra $\sB(\sH)$ is unitarily equivalent to the {\em
  amplification map} $\mapi{A}{A \tp \idty_\sE}$ for some Hilbert
space $\sE$ (\cite{sak}, Sect.~2.7). Any minimal Stinespring dilation of $T$ that has the
form (\ref{eq:stican}) will be referred to as its {\em canonical
  Stinespring dilation}. The canonical Stinespring dilation is
likewise unique up to unitary equivalence.

Another important structure theorem for CP maps is due to Kraus
\cite{kra}. It says that for any CP map $\map{T}{\sA}{\sB(\sH)}$, with
$\sA$ being a W*-algebra of operators on some Hilbert space $\sH'$,
there exists a collection of bounded operators $\map{V_x}{\sH}{\sH'}$,
such that
\begin{equation}
T(A) = \sum_x V^*_x A V_x,
\label{eq:krausform}
\end{equation}
where the series converges in the strong operator topology. If $\dim
\sH = \infty$, the set $\set{V_x}$ can be chosen in such a way that
its cardinality equals the Hilbertian dimension (i.e., the cardinality
of any complete orthonormal basis) of $\sH$ \cite{el}.

The Stinespring dilation (\ref{eq:stican}) and the Kraus form
(\ref{eq:krausform}) of a CP map $\map{T}{\sB(\sH_1)}{\sB(\sH_2)}$ are
related to one another via the correspondence
\begin{equation}
V\psi = \sum_x V_x \psi \tp e_x \qquad \all \psi \in \sH_2,
\label{eq:kracan}
\end{equation}
where $\set{e_x}$ is an orthonormal system in $\sE$. Note that the
Kraus operators $\set{V_x}$ depend on the choice of $\set{e_x}$. The
adjoint operator $\map{V^*}{\sH_1 \tp \sE}{\sH_2}$ acts on the
elementary tensors $\psi \tp \chi \in \sH_1 \tp \sE$ as
$$
V^*(\psi \tp \chi) = \sum_x \braket{e_x}{\chi}V^*_x\psi.
$$
It is not hard to see that when $\sH_1$ and $\sH_2$ are both
finite-dimensional, any canonical Stinespring dilation of $T$ will
give rise to at most $\dim \sH_1 \cdot \dim \sH_2$ Kraus
operators. This is so because these Kraus operators must be linearly
independent elements of the vector space $\sL(\sH_2,\sH_1)$ of all linear
operators from $\sH_2$ into $\sH_1$. Furthermore, the number
of terms in such a Kraus decomposition is uniquely determined by $T$ \cite{ls}.

\paragraph{Partial order of CP maps ---} The cone $\cp(\sA;\sH)$ of all
normal CP maps of $\sA$ into $\sB(\sH)$ can be partially ordered in
the following natural fashion. Given $S,T \in \cp(\sA;\sH)$, we will
write $S \le T$ if $T - S \in \cp(\sA;\sH)$. Following Belavkin and
Staszewski \cite{bs}, we will say that $S$ is {\em completely
  dominated by $T$}. Given a nonnegative real constant $c$, we will
say that $S$ is {\em completely $c$-dominated by $T$} if $S \le cT$. Using
the condition (\ref{eq:cpcond}), we see that $S \le T$ if and only if
$$
\sum^n_{i,j=1}\braket{\eta_i}{S(A^*_iA_j)\eta_j} \le
\sum^n_{i,j=1}\braket{\eta_i}{T(A^*_iA_j)\eta_j} \qquad \all \eta_i
\in \sH, A_i \in \sA; i =1,\ldots, n
$$
for each $n \in \N$. We will use the notation $\cp(\sH_1,\sH_2)$ (note
the comma) for the set of all CP maps of $\sB(\sH_1)$ into
$\sB(\sH_2)$.

\subsection{Quantum operations}
\label{ssec:qops}

Reversible dynamics of a closed
quantum-mechanical system
with the Hilbert space $\sH$ is given, in the \Schrodinger\ picture,
by the mapping $\mapi{\rho}{U\rho U^*}$, where $\rho$ is a density
operator on $\sH$ (i.e., $\tr {\rho} = 1$ and $\rho \ge 0$), and
$\map{U}{\sH}{\sH}$ is a unitary transformation. In the dual
Heisenberg picture the same dynamics is described by the mapping
$\mapi{A}{U^*AU}$ for all $A \in \sB(\sH)$.  The two descriptions are
equivalent as they yield the same observed statistics, $\tr(U\rho
U^*A) = \tr(\rho U^*A U)$. 

On the other hand, when the system is open because it is either
coupled to an environment or is being subjected to a measurement, its
most general time evolution is irreversible.  This is captured
mathematically by means of a {\em quantum operation} \cite{kra}, i.e., a
completely positive normal linear map $\map{T}{\sB(\sH)}{\sB(\sH)}$ with the
additional constraint $T(\idty) \le \idty$.  In terms of the Kraus
form, $T(A) = \sum_x V^*_x A V_x$, we have the bound $\sum_x V^*_x V_x \le
\idty$.  The corresponding \Schrodinger-picture map on density
operators, $\mapi{\rho}{T_*(\rho)}$, is defined \cite{note1} by
$$
\tr[T_*(\rho)A] =
\tr[\rho T(A)] \qquad \all A \in \sB(\sH),
$$
and can then be extended to the linear span of the density operators,
the trace class $\sT(\sH)$. It follows at once that the map $T_*$ is
completely positive and trace-decreasing in the sense that $\tr
T_*(X) \le \tr X$ for any $X \in \sT(\sH)$.  In order to retain proper
normalization for density operators, one usually writes the
\Schrodinger-picture evolution dual to $T$ as
$\mapi{\rho}{T_*(\rho)/\tr T_*(\rho)}$.  Alternatively, one says that
the transformation $\mapi{\rho}{T_*(\rho)}$ {\em succeeds with
  probability $\tr T_*(\rho)$}; this probability is equal to unity for
all density operators $\rho$ if and only if $T$ is unital, i.e.,
$T(\idty) = \idty$, so that $T_*$ is trace-preserving. Unital quantum
operations are also referred to as {\em quantum channels} \cite{key}.

The Kraus theorem implies that we can write any quantum operation $T$
as a sum of {\em pure operations} (\cite{dav}, Sect.~2.3), i.e., maps of the form
$\mapi{A}{X^*AX}$ with $X^*X \le \idty$ (this is equivalent to $X$
being a contraction, $\norm{X} \le 1$ where $\norm{\cdot}$ is the
usual operator norm, $\norm{X} = \sup_{\psi \in
  \sH}\norm{X\psi}/\norm{\psi}$).  The qualification ``pure'' is
usually interpreted as referring to
the fact that, for any pure state $\ketbra{\psi}{\psi}$, the
(unnormalized) state $X\ketbra{\psi}{\psi}X^*$ is pure as
well \cite{dpp}. However, as we shall see later, it is a direct
consequence of the Radon-Nikodym theorem for CP maps that $T$ is a
pure operation if and only if all operations completely dominated by
it are its nonnegative multiples.  This is analogous to the case of
pure states on a C*-algebra $\sA$:  a state $\omega$ on $\sA$ is pure
if and only if all positive linear functionals $\varphi$ on $\sA$,
such that $\omega - \varphi$ is positive are nonnegative multiples of
$\omega$ (\cite{br}, Sect.~2.3.2).

Given the canonical Stinespring dilation (\ref{eq:stican})
of a quantum channel $T$ (in which case $V$ is an isometry), the
\Schrodinger-picture operation $T_*$ can be cast in the so-called {\em
  ancilla form}
\begin{equation}
T_*(\rho) = \tr_\sE U(\rho \tp \ketbra{\xi}{\xi})U^*,
\label{eq:ancilla}
\end{equation}
where $\tr_\sE (\cdot)$ denotes the partial trace over $\sE$, $\xi \in
\sE$ is a fixed unit vector, and $U$ is the unitary extension of the
partial isometry $\hU$ from $\sH_2 \tp [\ketbra{\xi}{\xi}]$ to $\sH_1
\tp \sE$ defined by $\hU (\psi \tp \xi) = V\psi$
\cite{kra,oza0}. (We use $[P]$ to denote the closed subspace corresponding to
the orthogonal projection $P$.)

Finally, note that the input and output Hilbert spaces do not have to be the
same; in general, quantum operations are completely positive normal
linear maps $\map{T}{\sB(\sH_1)}{\sB(\sH_2)}$ with $T(\idty_{\sH_1})
\le \idty_{\sH_2}$.  The corresponding \Schrodinger-picture operations are
completely positive trace-decreasing maps
$\map{T_*}{\sT(\sH_2)}{\sT(\sH_1)}$.  Most of the discussion in this
section carries over to this case, modulo straightforward
modifications; however, one must be careful with the ancilla
representation of a general \Schrodinger-picture channel $T_*$.  The
key caveat here is that the initial ancillary space and the final
``traced-out'' space need not be isomorphic.  This yet again
underscores the advantages of working in the Heisenberg picture.

\subsection{The norm of complete boundedness}
\label{ssec:cbnorm}

In many information-theoretic studies of noisy quantum channels one
needs a quantitative measure of the ``noisiness'' of a channel; this
is, in fact, a natural departure point for various definitions of
information-carrying capacities of quantum channels
\cite{key,hw,wer}. A good candidate for such a measure is the norm
$\norm{T-\id}_?$, where the question mark refers to the fact that we
have not yet specified a suitable norm.

The choice of the proper norm turns out to be a tricky matter \cite{key}. Let
$\sA$ and $\sB$ be C*-algebras, and consider a linear map
$\map{\Lambda}{\sA}{\sB}$. We cannot adopt the operator norm, defined by
\begin{equation}
\norm{\Lambda} = \sup \setcond{\norm{\Lambda(A)}}{A \in \sA, \norm{A} \le 1},
\label{eq:opnorm}
\end{equation}
where $\norm{A}$ is the (unique) C*-norm on $\sA$, because the norm
  $\norm{\Lambda \tp \id_n}$ of the map $\map{\Lambda\tp\id_n}{\sA \tp
  \sM_n}{\sB\tp\sM_n}$ can increase with $n$ even if $\Lambda$ itself is
  bounded (see Ch.~3 of \cite{pis}). What
  we need is a ``stabilized'' version of (\ref{eq:opnorm}). A map
$\map{\Lambda}{\sA}{\sB}$ is called {\em completely bounded} (CB for short) if there exists
some constant $C \ge 0$ such that all the maps $\map{\Lambda \tp \id_n}{\sA
  \tp \sM_n}{\sB \tp \sM_n}$ are uniformly bounded by $C$, i.e., $\norm{\Lambda
  \tp \id_n} \le C$. The CB norm $\cbnorm{\Lambda}$ is defined to be the
smallest constant $C$ for which this holds, i.e.,
$$
\cbnorm{\Lambda} = \sup_{n \in \N}\norm{\Lambda \tp \id_n}.
$$
All CB maps have the
property of ``factoring through a Hilbert space,'' as shown in the
following key structure theorem (Thm.~3.6 in \cite{pis}), given here in a
slightly simplified form suitable for our needs.

\begin{theorem}[Haagerup-Paulsen-Wittstock]\label{th:hpw} Let $\sH$
  and $\sK$ be Hilbert spaces, and let
  $\map{\Lambda}{\sB(\sH)}{\sB(\sK)}$ be a CB map. Then there exist a
  Hilbert space $\sE$ and operators $\map{V_1,V_2}{\sK}{\sH \tp \sE}$
  with $\norm{V_1} \norm{V_2} \le \cbnorm{\Lambda}$ ($\norm{\cdot}$
  stands for the operator norm), such that
\begin{equation}
\Lambda(A) = V^*_1 (A
  \tp \idty_\sE)V_2. 
\label{eq:cbfactor}
\end{equation}
Conversely, any map $\Lambda$ of the form (\ref{eq:cbfactor}) satisfies
$\cbnorm{\Lambda} \le \norm{V_1}\norm{V_2}$.
\end{theorem}

\noindent{Note that the Stinespring and the Haagerup-Paulsen-Wittstock
theorems together imply that any CP map is automatically CB. In fact,
for a CP map $T$, we have $\cbnorm{T} = \norm{T(\idty)}$ \cite{pau}. Also, the
difference of two CP maps is always CB.}

Theorem \ref{th:hpw} suggests an alternative way to define the CB norm of a
map $\Lambda$, namely as
\begin{equation}
\cbnorm{\Lambda} = \inf \set{\norm{V_1}\norm{V_2}},
\label{eq:altcb}
\end{equation}
where the infimum is taken over all possible decompositions of
$\Lambda$ in the form (\ref{eq:cbfactor}). Moreover, the theorem
guarantees that the infimum in (\ref{eq:altcb}) is attained.

In quantum information theory one frequently deals with both the
operation $\map{T}{\sB(\sH)}{\sB(\sK)}$ and its (pre)dual,
$\map{T_*}{\sT(\sK)}{\sT(\sH)}$. As we mentioned in
Sect.~\ref{ssec:qops}, $T$ and $T_*$ are connected by the relation
$\tr[T(A)B] = \tr[AT_*(B)]$, $A \in \sB(\sH)$, $B \in \sT(\sK)$. This
duality holds also for any normal CB map
$\map{\Lambda}{\sB(\sH)}{\sB(\sK)}$, so that when $\Lambda$ is written
in the form (\ref{eq:cbfactor}), we have
\begin{equation}
\Lambda_*(A) = \tr_\sE V_2 A V^*_1 \qquad \all A \in \sT(\sK).
\label{eq:cbfactordual}
\end{equation}
This motivates the definition of the dual CB norm,
\begin{equation}
\cbnorm{\Lambda_*}^* = \inf \set{\norm{V_1}\norm{V_2}},
\label{eq:cbnormdual}
\end{equation}
where the infimum
is taken over all possible decompositions of $\Lambda_*$ in the form
(\ref{eq:cbfactordual}). It is now clear that $\cbnorm{\Lambda} =
\cbnorm{\Lambda_*}^*$ for any normal CB map $\Lambda$, so in the
future we will always write $\cbnorm{\Lambda}$, even when working with
$\Lambda_*$. In fact, the norm (\ref{eq:cbnormdual}) was introduced
by Kitaev \cite{kit} under the name ``diamond norm'' (Kitaev used the
notation $\dnorm{\Lambda}$). The equivalence of the diamond norm and
the CB norm has been alluded to in the literature on quantum
information theory \cite{hw} but, to the best of our knowledge, no
proof of the equivalence was ever presented. 

The duality relation between $\map{\Lambda}{\sB(\sH)}{\sB(\sK)}$ and
$\map{\Lambda_*}{\sT(\sK)}{\sT(\sH)}$ implies that we can also write
$$
\cbnorm{\Lambda} = \sup_{n \in \N} \trnorm{\Lambda_* \tp \id_n},
$$ where $\trnorm{\Lambda_*} = \sup\setcond{\trnorm{\Lambda_*(A)}}{A \in
  \sT(\sK), \trnorm{A} \le 1}$ and $\trnorm{A} = \tr \abs{A} \equiv \tr
  \sqrt{A^*A}$ is the trace norm (Sect.~VI.6 \cite{rs}). For this purpose we can
  use the well-known variational characterization of the operator norm
  (Thm.~3.2 in \cite{sim}), namely
$$
\norm{A} = \sup_{B \in \sT(\sH) \atop \trnorm{B} \le 1} \abs{\tr (AB)}
\qquad \all A \in \sB(\sH).
$$
Then for any normal CB map
  $\map{\Lambda}{\sB(\sH)}{\sB(\sK)}$ we have
\begin{eqnarray*}
\norm{\Lambda} &=& \sup_{A \in \sB(\sH) \atop \norm{A} \le 1}
\norm{\Lambda(A)} = \sup_{B \in \sT(\sK) \atop \trnorm{B} \le 1}
\sup_{A \in \sB(\sH) \atop \norm{A} \le 1} \abs{\tr [\Lambda(A)B]} \\
&=&  \sup_{B \in \sT(\sK) \atop \trnorm{B} \le 1}
\sup_{A \in \sB(\sH) \atop \norm{A} \le 1} \abs{\tr [A\Lambda_*(B)]}
= \sup_{B \in \sT(\sK) \atop \trnorm{B} \le 1} \trnorm{\Lambda_*(B)}
= \trnorm{\Lambda_*},
\end{eqnarray*}
which also implies that $\norm{\Lambda \tp \id_n} = \trnorm{\Lambda_*
  \tp \id_n}$ for all $n \in \N$. Taking the supremum of both sides
  with respect to $n$ does the job. In a nutshell, the CB norm of a map
  between algebras of bounded operators on Hilbert spaces can be
  defined through a variational expression involving the operator
  norm, whereas the CB norm of the corresponding dual map between the
  trace classes is determined by a variational expression in the trace norm.

We now summarize the key properties of the CB norm. For any two CB
maps $\map{\Lambda}{\sB(\sH)}{\sB(\sH')}$ and
$\map{\Lambda}{\sB(\sH')}{\sB(\sK')}$, any $A \in \sB(\sH)$, and any
$B \in \sT(\sH')$, we have the following.
\begin{enumerate}
\item $\cbnorm{\Lambda' \circ \Lambda} \le
  \cbnorm{\Lambda'}\cbnorm{\Lambda}$;
\item $\cbnorm{\Lambda \tp \Lambda'} =
  \cbnorm{\Lambda}\cbnorm{\Lambda'}$;
\item $\norm{\Lambda(A)} \le \cbnorm{\Lambda}\norm{A}$;
\item $\trnorm{\Lambda_*(B)} \le \cbnorm{\Lambda}\trnorm{B}$.
\end{enumerate}
For proofs see, e.g., the article of Kitaev \cite{kit} or the
monographs of Pisier \cite{pis} and Paulsen \cite{pau}.

\section{The Radon-Nikodym theorem for completely positive maps}
\label{sec:rnt}

In this section we review a theorem of the Radon-Nikodym type that
allows for a complete classification of all CP maps $S$ that are
completely dominated by a given CP map $T$. As we have already
mentioned, this theorem can be distilled from the more general results of
Arveson \cite{arv} and Belavkin and Staszewski \cite{bs}. The
work of Parthasarathy \cite{par} contains further developments, in
particular an analogue of the Lebesgue decomposition for CP maps. The
idea is to express all maps $S$ that satisfy $S \le T$ in the form
related to the (minimal) Stinespring dilation of $T$; this
``Stinespring form'' of the theorem \cite{arv,bs} is stated in
Sect.~\ref{ssec:rnsti}, with the proof included in order to keep the
paper self-contained. Then, in Sect.~\ref{ssec:rnkra}, we state and
prove two ``Kraus forms'' of the Radon-Nikodym theorem. Finally,
some general remarks are given in Sect.~\ref{ssec:rngen}.

\subsection{The Stinespring form}
\label{ssec:rnsti}

Before we state and prove the Radon-Nikodym theorem, let us recall a
standard piece of notation.  Given a C*-algebra $\sA$ and a
$*$-homomorphism $\map{\pi}{\sA}{\sB(\sH)}$, the set $\setcond{B \in
  \sB(\sH)}{[A,B]\equiv AB-BA=0, \all A \in \pi(\sA)}$ is called the {\em
  commutant of $\pi$} and is denoted by $\pi(\sA)'$.

\begin{theorem}\label{th:rn} Consider $S,T \in \cp(\sA;\sH)$, and
  let $(\sK,V,\pi)$ be the minimal Stinespring dilation of
  $T$. Then $S \le T$ if and only if there exists an operator $\hF \in
  \pi(\sA)'$, such that $0 \le \hF \le \idty$ and
$$
S(A) = V^*\pi(A)\hF V = V^* \hF^{1/2}\pi(A)\hF^{1/2}V
$$
for all $A \in \sA$. The operator $\hF$ is unique in the sense that if
$S(A) = V^*\pi(A)YV$ for some $Y \in \pi(\sA)'$,  then $Y = \hF$. We
will refer to this operator $\hF$ as the {\em Radon-Nikodym derivative
  of $S$ with respect to $T$} and denote it by $\dd_TS$.

\end{theorem}

\begin{proof} Suppose $S \le T$, and let $(\sK',V',\pi')$ be the
  minimal Stinespring dilation of $S$. Define an operator
  $\map{\hG}{\sK}{\sK'}$ by
$$
\imap{\hG}{\pi(A)V\eta}{\pi'(A)V'\eta} \qquad \all A \in \sA, \eta \in \sH,
$$
and extend it to the linear span of $\pi(\sA)V\sH$. For any finite
linear combination $\Psi = \sum^n_{i=1}\pi(A_i)V\eta_i$ we have
\begin{eqnarray*}
\norm{\hG\Psi}^2 &=& \sum^n_{i,j=1}\braket{\eta_i}{V'^*\pi'(A^*_iA_j)V'\eta_j}
= 
\sum^n_{i,j=1}\braket{\eta_i}{S(A^*_iA_j)\eta_j} \\
&\le&
\sum^n_{i,j=1}\braket{\eta_i}{T(A^*_iA_j)\eta_j} =
\sum^n_{i,j=1}\braket{\eta_i}{V^*\pi(A^*_i A_j)V\eta_j} =  \norm{\Psi}^2.
\end{eqnarray*}
Thus $\hG$ is a densely defined contraction, and
therefore extends to a contraction from $\sK$ into $\sK'$. We will
denote this extension also by $\hG$. For the adjoint map $\hG^*$, we
have
$$
\braket{\eta}{V^*\hG^*\pi'(A)V'\xi} = \braket{\hG V
  \eta}{\pi'(A)V'\xi} = \braket{V'\eta}{\pi'(A)V'\xi} =
\braket{\eta}{V'^*\pi'(A)V'\xi} \equiv \braket{\eta}{S(A)\xi}
$$
for all $\eta,\xi \in \sH$ and $A \in \sA$, which implies that
$V^*\hG^*\pi'(A)V'\eta = S(A)\eta$.

The map $\hG$ intertwines the representations $\pi$ and $\pi'$, i.e.,
$\hG \pi(A) = \pi'(A)\hG$ for any $A \in \sA$. Indeed, for all $A, B \in
\sA$ and $\eta \in \sH$ we have
$$
\hG\pi(A)\pi(B)V \eta = \hG\pi(AB)V\eta = \pi'(AB)V'\eta =
\pi'(A)\pi'(B)V'\eta = \pi'(A)\hG \pi(B)V\eta,
$$
and the desired statement follows because of the minimality of the
Stinespring dilation $(\sK,V,\pi)$. Taking adjoints, we also
obtain $\pi(A)\hG^* = \hG^*\pi'(A)$. Letting $\hF = \hG^*\hG$, we see
that
$$
\hF\pi(A) = \hG^*\hG\pi(A) = \hG^*\pi'(A)\hG = \pi(A)\hG^*\hG =
\pi(A)\hF,
$$
which shows that $\hF \in \pi(\sA)'$. Finally, for all $A \in \sA$ and
$\eta \in \sH$ we have
$$
V^*\hF \pi(A) V\eta = V^*\hG^*\hG\pi(A)V\eta = V^*\hG^*\pi'(A)V'\eta =
S(A)\eta,
$$
thus $S(A) = V^*\hF\pi(A)V = V^*\pi(A)\hF V =
V^*\hF^{1/2}\pi(A)\hF^{1/2}V$. The uniqueness of $\hF$ follows from
the minimality of $(\sK',V',\pi')$.

The converse is clear.
\end{proof}

For the special case $S,T \in \cp(\sH_1,\sH_2)$ we can use the
canonical Stinespring dilation (\ref{eq:stican}) and the fact
that the commutant of the algebra $\sB(\sH_1) \tp \C \idty_\sE$ is
isomorphic to $\C \idty_{\sH_1} \tp \sB(\sE)$ (Thm.~IV.5.9 in \cite{tak}),
to deduce the following.

\begin{corollary}\label{cor:rn} Let $S,T \in \cp(\sH_1,\sH_2)$, and
  let $T(A) = V^*(A \tp \idty_\sE)V$ be the canonical Stinespring
  dilation of $T$. Then $S \le T$ if and only if there exists a
  positive contraction $F \in \sB(\sE)$, such that $S(A) = V^*(A \tp
  F)V$ for all $A \in \sB(\sH_1)$.\end{corollary}

As we already mentioned, the Radon-Nikodym theorem allows one to
fully appreciate the term ``pure operation.'' Let $\sH_1$ and $\sH_2$
be Hilbert spaces, and consider the map $T(A) = X^*AX$, where
$\map{X}{\sH_2}{\sH_1}$ is a contraction. Clearly, $X^*AX$ is the
canonical Stinespring dilation of $T$ so, by Theorem \ref{th:rn},
any $S \in \cp(\sH_1,\sH_2)$ that satisfies $S \le T$ must be of the
form $\lambda X^*AX$ for some $\lambda \in [0,1]$.

Theorem \ref{th:rn} can also be used to characterize completely all
ways to write a given $T \in \cp(\sA;\sH)$ as a finite sum $\sum_i T_i$,
with $T_i \in \cp(\sA;\sH)$ for all $i$. It is actually the resulting theorem,
stated below, that is referred to as the ``Radon-Nikodym theorem for
CP maps'' in the quantum information literature \cite{wer}.

\begin{theorem}\label{th:rnsum} Consider a map $T \in
  \cp(\sA;\sH)$ with the canonical Stinespring dilation $(\sK,V,\pi)$. For any finite decomposition $T = \sum_i T_i$ with
  $T_i \in \cp(\sA;\sH)$ there exist unique positive operators $\hF_i \in
  \pi(\sA)'$ that satisfy $\sum_i \hF_i = \idty_\sK$, such that $T_i(A) =
  V^*\pi(A)\hF_iV$.
\end{theorem}

\begin{proof} Apply Theorem \ref{th:rn} separately to each pair
  $(T_i,T)$, and let $\hF_i = \dd_TT_i$. Then $T(A) = \sum_i
  V^*\pi(A)\hF_i V = V^*\pi(A)V$, and $\sum_i \hF_i = \idty_\sK$ by
  the uniqueness part of Theorem \ref{th:rn}.
\end{proof}

\begin{rem}The decomposition $T=\sum_i T_i$ is a particularly simple
  instance of a CP instrument \cite{dav}. As such, it is not difficult
  to extract Theorem \ref{th:rnsum} from more general results of Ozawa
  \cite{oza}.
\end{rem}

\subsection{The Kraus form}
\label{ssec:rnkra}

Theorem \ref{th:rn} can be restated in a simple way in terms of the
Kraus form of a CP map. In order to do this, we need some additional
machinery (Sect.~II.15 in \cite{par1}).

Let $X$ be a set. Any function $\map{K}{X \times X}{\C}$ is called a
{\em kernel on $X$}. The set $\sK(X)$ of all kernels on $X$ is a vector space, with the corresponding algebraic operations
defined pointwise on $X \times X$. We say that a
kernel $K \in \sK(X)$ is {\em positive-definite}, and write $K \ge 0$,
if for each $n \in \N$ we have
$$
\sum^n_{i,j=1} \overline{c_i}{c_j} K(x_i,x_j)\ge 0 \qquad \all x_i \in
X, c_i \in \C; i = 1,\ldots,n.
$$
Given a pair of kernels $K,K' \in \sK(X)$, we will write $K \le K'$ if
$K' - K$ is positive-definite. Note that a positive-definite kernel is
automatically Hermitian, i.e., $K(x,y) = \overline{K(y,x)}$.

According to the fundamental theorem of Kolmogorov, for any
positive-definite kernel $K \in \sK(X)$ there exist a Hilbert space
$\sH_K$ and a map $\map{v_K}{X}{\sH_K}$ such that
$\braket{v_K(x)}{v_K(y)} = K(x,y)$ for all $x,y \in X$, and the set
$\setcond{v_K(x)}{x \in X}$ is total in $\sH_K$. The pair
$(\sH_K,v_K)$ is referred to as the {\em Kolmogorov decomposition of
  $\sK$} and is unique up to unitary equivalence.

After these preparations, we may state our first result.

\begin{theorem}\label{th:rnkraus} Consider two maps $S, T \in
  \cp(\sH_1,\sH_2)$.  Let $\set{V_x}_{x \in X}$ be a Kraus
  decomposition of $T$ induced by the canonical Stinespring
  dilation $T(A) = V^*(A \tp \idty_\cE)V$, as prescribed in
  (\ref{eq:kracan}). Then $S \le T$ if and only if
$$
S(A) = \sum_{x,y \in X}K(x,y)V^*_x A V_y
$$
for some positive-definite kernel $K \in \sK(X)$ with $K \le I$,
where $I$ is the Kronecker kernel $I(x,y) \equiv \delta_{xy}$.
\end{theorem}

\begin{proof} Suppose $S \le T$. By Corollary \ref{cor:rn},
  $S(A) = V^*(A \tp F)V$ for some positive contraction $F \in
  \sB(\sE)$. Let $\set{e_x}_{x \in X}$ be the orthonormal system in $\sE$,
  determined by $V$ and $\set{V_x}$ from (\ref{eq:kracan}). Then for
  any $\eta \in \sH_2$ we have
$$
S(A)\eta = V^*(A \tp F)V\eta = V^*\left(\sum_{y \in X}AV_y\eta \tp
Fe_y\right) = \sum_{x,y \in X} \braket{e_x}{Fe_y}V^*_x A V_y \eta.
$$
Define the kernel $K \in \sK(X)$ by setting $K(x,y) \defeq
\braket{e_x}{Fe_y}$. Then $0 \le F \le \idty$ implies that $0 \le K
\le I$.

Conversely, suppose we are given
$$
T(A) = \sum_{x \in X} V^*_x A V_x
$$
and
$$
S(A) = \sum_{x,y \in X}K(x,y)V^*_x A V_y
$$
for some $K \in \sK(X)$ such that $0 \le K\le I$. Let $(\sH_K,v_K)$ be
the Kolmogorov decomposition of $K$, and let ${\rm fin}(X)$ be the set of all
finite subsets of $X$. Define an operator $\map{G}{\sE}{\sH_K}$ by
$$
\imap{G}{\sum_{x \in X_0} c_x e_x}{\sum_{x \in X_0} c_x v_K(x)}
  \qquad \all c_x \in \C, X_0 \in {\rm fin}(X).
$$
It is easy to see that, for any $X_0 \in {\rm fin}(X)$,
$$
\Norm{\sum_{x \in X_0}c_x e_x}^2 = \sum_{x \in X_0} \abs{c_x}^2 = 0
$$
implies $c_x = 0$ for all $x \in X_0$, and consequently
$$
\Norm{G\left(\sum_{x \in X_0}c_x e_x\right)}^2 = \sum_{x,y \in X_0}
\overline{c_x}c_y K(x,y) \le \sum_{x \in X_0} \abs{c_x}^2 = 0,
$$
where the last equality above follows because $K \le I$. Thus $G$
extends to a well-defined linear operator on $\sE$, which we will also
denote by $G$. Let $F =
G^*G$. Then $\braket{e_x}{Fe_y} = \braket{v_K(x)}{v_K(y)} = K(x,y)$,
and $0 \le K \le I$ implies that $0 \le F \le \idty_\sE$. Thus, for
all $A \in \sB(\sH_1)$ and $\eta \in \sH_2$ we have
$$
S(A)\eta = \sum_{x,y}K(x,y)V^*_x A V_y = \sum_{x,y \in
  X}\braket{e_x}{Fe_y} V^*_x A V_y = V^*(A \tp F)V,
$$
so that $S \le T$ by Corollary \ref{cor:rn}.
\end{proof}

\begin{rems} 1. When the set $\set{V_x}$ is finite, Theorem
  \ref{th:rnkraus} says that $S \le T$ for $T(A) = \sum_x V^*_x A V_x$
  if and only if $S(A) = \sum_{x,y} M_{xy}V^*_x A V_y$ for some matrix
  $M = [M_{xy}]$ with $0 \le M \le \idty$.

\noindent{2. Since we deal only with separable Hilbert spaces, the
  index set $X$ is at most countably infinite.}
\end{rems}

Another Kraus form of the Radon-Nikodym theorem can be
proved directly, without recourse to the theory of positive-definite
kernels.

\begin{theorem}\label{th:rnkraus2} Consider two maps $S,T \in
  \cp(\sH_1,\sH_2)$. Then $S \le T$ if and only if there exist a
  Kraus decomposition $T(A) = \sum_x W^*_x A W_x$, induced by the
  canonical Stinespring dilation of $T$, and a set
  $\setcond{\lambda_x}{\lambda_x \in [0,1]}$, such that $S(A) =
  \sum_x \lambda_x W^*_x A W_x$.
\end{theorem}

\begin{proof} Suppose $S \le T$. Let $T(A) = V^*(A \tp \idty_\sE)V$ be
  the canonical Stinespring dilation of $T$. Then Corollary
  \ref{cor:rn} says that $S(A) = V^*(A \tp F)V$ for some positive
  contraction $F \in \sB(\sE)$. Write down the spectral decomposition
  $F = \sum_x \lambda_x \ketbra{\phi_x}{\phi_x}$, so that $\lambda_x
  \in [0,1]$ and $\braket{\phi_x}{\phi_y}=\delta_{xy}$. Let $\set{W_x}$ be
  the Kraus decomposition of $T$ determined from (\ref{eq:kracan}) by
  $V$ and $\set{\phi_x}$. Then for any $\eta \in \sH_2$ we have
\begin{eqnarray*}
S(A)\eta &=& V^*(A \tp F)V\eta
= V^*\left(\sum_y \lambda_y AW_y \eta \tp \phi_y\right) \\
&=&
\sum_{x,y}\lambda_y \braket{\phi_x}{\phi_y} W^*_x A W_y \eta = \sum_x
\lambda_x W^*_x A W_x.
\end{eqnarray*}
The converse follows readily from the fact that the map
$\mapi{A}{\sum_x (1-\lambda_x)W^*_x A W_x}$ is CP for any choice of
$\set{W_x}$ and $\set{\lambda_x}$ with $\lambda_x \in [0,1]$.
\end{proof}

\subsection{General remarks}
\label{ssec:rngen}

Before we go on, we would like to
pause and make some general comments about the significance of the
Radon-Nikodym theorem for CP maps at large.

The real power of this theorem lies in the fact that it contains the
``traditional'' forms of the Radon-Nikodym theorem as special
cases.  In order to see this, we will need the following result (see
Corollary IV.3.5 and Proposition IV.3.9 in \cite{tak}):  a positive
map $T$ from a C*-algebra $\sA$ to another C*-algebra $\sB$ is
automatically completely positive whenever at least one of $\sA$ and
$\sB$ is Abelian.    

With this in mind, let us observe that any positive linear functional
$\varphi$ on a C*-algebra $\sA$ is a positive map from $\sA$ to $\C$, and
therefore is CP.  When we apply the Stinespring theorem to $\varphi$,
we simply recover the GNS representations $(\sH,\pi,\Omega)$ of $\sA$
induced by $\varphi$, where $\sH$ is
the Hilbert space of the representation, $\pi$ is a $*$-isomorphism
between $\sA$ and a suitable C*-subalgebra of $\sB(\sH)$, and
$\Omega \in \sH$ is cyclic for $\pi$, i.e., $\sH =
\overline{\pi(\sA)\Omega}$. Of course, we have then $\varphi(A) =
\braket{\Omega}{\pi(A)\Omega}$ for all $A \in \sA$.

Consider first the Abelian case.  Let $X$ be a compact Hausdorff
space, and let $\sA$ be the commutative C*-algebra $\sC(X)$ of all complex-valued
continuous functions on $X$.  Let $\varphi$ be a positive linear
functional on $\sC(X)$.  By the Riesz-Markov theorem (see Thm.~IV.14 \cite{rs}), there exists a unique Baire measure $\mu$ on $X$ such
that $\varphi(f) = \int_X f(x)d\mu(x), \all f \in \sC(X)$. If
$\varphi$ is a state [i.e., $\varphi(\idty_X) = 1$ where $\idty_X$ is, of
course, the function on $X$ that is identically equal to 1], then $\mu$ is
a probability measure. The GNS construction yields the cyclic
representation $(\sH,\pi,\Omega)$, where $\sH = \sL^2(X,d\mu)$,
$[\pi(f)g](x) = f(x)g(x)$, and $\Omega = \idty_X$, such that
$$
\varphi(f) = \braket{\Omega}{\pi(f)\Omega} = \int_X f(x)d\mu(x).
$$
This is the minimal Stinespring dilation of the CP map
$\map{\varphi}{\sC(X)}{\C}$; more precisely, we have the isometry
$\map{V}{\C}{\sL^2(X,d\mu)}$ defined by $Vc = c\Omega$, so that
  $\varphi(f) = V^*\pi(f)V$. Now suppose we are given another positive
linear functional $\eta$ on $\sC(X)$ such that $\eta \le \varphi$,
i.e., $\eta(f) \le \varphi(f)$ for every nonnegative $f \in
\sC(X)$.  Then Theorem \ref{th:rn} states that there exists a
nonnegative function $\rho \in \pi(\sC(X))' \subseteq
\sL^\infty(X,d\mu)$ such that $\eta(f) = V^*\pi(f)\rho V$, i.e., 
$$
\eta(f) = \braket{\Omega}{\rho \pi(f)\Omega} = \int_X
\rho(x)f(x)d\mu(x).
$$
Again, by the Riesz-Markov theorem, there exists a unique Baire measure $\nu$
on $X$ such that $\eta(f) = \int_X f(x)d\nu(x)$.  It is easy to see
that the function $\rho$ is precisely the measure-theoretic
Radon-Nikodym derivative $d\nu / d\mu$.

The noncommutative case is dealt with in a similar manner.
Namely, if $\varphi$ is a state on a unital C*-algebra $\sA$ that
admits the cyclic representation $(\sH,\pi,\Omega)$, then any positive
linear functional $\eta$ on $\sA$ such that $\eta \le \varphi$ has the
form $\eta(A) = \braket{\Omega}{\pi(A)F\Omega}$ for a unique positive
contraction $F \in \pi(\sA)'$.  This is, of course, the familiar
Radon-Nikodym theorem for states on C*-algebras (see Thm.~2.3.19 in \cite{br}).

\section{Partial ordering of quantum operations}
\label{sec:poqo}

The first series of problems we tackle by means of the Radon-Nikodym
theorems of Sect.~\ref{sec:rnt} is connected to the partial ordering of quantum
operations with respect to the relation of complete domination, defined in
Sect.~\ref{sec:prelims}. 

As mentioned already, all quantum operations
$\map{T}{\sB(\sH_1)}{\sB(\sH_2)}$ must satisfy $T(\idty_{\sH_1}) \le
\idty_{\sH_2}$. It turns out that this normalization condition imposes
severe restrictions on the structure of their order intervals.  In
particular, as shown in the following Proposition, no nontrivial
difference of quantum channels can be a CP map.

\begin{proposition}\label{prop:chdif} Let $S,T \in \cp(\sH_1,\sH_2)$ be quantum
  channels.  Then $T-S \in \cp(\sH_1,\sH_2)$ if and only if $S = T$.
\end{proposition}

\begin{proof} Suppose $T - S \in \cp(\sH_1,\sH_2)$, or, equivalently,
  $S \le T$. Then Theorem \ref{th:rnkraus2} implies that there exists a
  Kraus decomposition $T(A) = \sum_x W^*_x A W_x$ such that $S(A) =
  \sum_x \lambda_x W^*_x A W_x$ with $0 \le \lambda_x \le 1$. Because
  both $S$ and $T$ are channels, $S(\idty) = T(\idty) = \idty$, which
  implies that $\sum_x (1-\lambda_x)W^*_x W_x = 0$. Since each term in
  this sum is a positive operator, the only possibility is that
  $\lambda_x = 1$ for all $x$, or $S = T$. The converse is obvious.
\end{proof}

\begin{rem} To obtain an even simpler proof of this proposition, we
  can use the fact that, for a CP map $T$, $\cbnorm{T} = \norm{T(\idty)}$
  (cf. Sect.~\ref{ssec:cbnorm}). Indeed, if $S$ and $T$ are channels,
  then $T(\idty)=S(\idty)=\idty$, and the assumption that $T-S$ is CP
  yields $\cbnorm{T-S}=\norm{T(\idty)-S(\idty)}=0$, or $S=T$. In fact,
  the same method shows that if $S$ and $T$ are two CP maps with
  $S(\idty)=T(\idty)$, then $S-T$ cannot be a CP map.
\end{rem}

The only possible order
relation between a pair of quantum channels $S$ and $T$ is that, say,
$T$ completely $c$-dominates $S$ for some $c > 1$. Tthe latter
condition follows from Proposition \ref{prop:chdif} and from the fact that $S
\le cT$ implies $\idty \le c\idty$, which is (trivially) possible only
if $c \ge 1$. In fact, as pointed out by Parthasarathy \cite{par}, there are
pairs of channels $T$, $T'$ for which there exist constants $c,c' > 1$
such that $T' \le cT$ and $T \le c'T'$. To show this, let $S_1$ and $S_2$
be arbitrary channels, and define $T = \lambda S_1 + (1-\lambda)S_2$
and $T' = \lambda' S_1 + (1-\lambda')S_2$, where $0 < \lambda,
\lambda' < 1$. Then, setting $c = [\lambda (1-\lambda)]^{-1}$ and $c'
= [\lambda' (1-\lambda')]^{-1}$, we see that indeed $T' \le cT$ and $T
\le c'T'$. In Parthasarathy's terminology \cite{par}, $T$ and $T'$ are
    {\em uniformly equivalent}; this is written $T \equiv_u T'$, and
    is an equivalence relation.

The next problem we consider has to do with an alternative way to
(partially) order quantum operations by means of orthogonal
projections on a suitably enlarged Hilbert space. To this end we need
to recall some facts about the so-called {\em positive operator-valued
  measures} (POVM's for short) (Sect.~3.1 in \cite{dav}). Let $X$ be a
topological space, $\Sigma_X$ the $\sigma$-algebra of all Borel
subsets of $X$, and $\sH$ a Hilbert space. A map
$\map{M}{\Sigma_X}{\sB(\sH)}$ is a POVM on (the Borel subsets of) $X$
if it has the following properties:
\begin{enumerate}
\item (normalization) $M(\emptyset) = 0$ and $M(X) = \idty$.
\item (positivity) $M(\Delta) \ge 0$ for all $\Delta \in \Sigma_X$.
\item ($\sigma$-additivity) If $\set{\Delta_i}$ is a countable
  collection of pairwise disjoint Borel sets in $X$, then
  $M\left(\bigcup_i \Delta_i\right) = \sum_i M(\Delta_i)$, where the
  sum converges in the weak operator topology.
\end{enumerate}
A POVM that satisfies an additional requirement that  each $M(\Delta)$
is an orthogonal projection, i.e., $M(\Delta)^2 = M(\Delta)$, is
called a {\em projection-valued measure} (PVM). The resulting
resolution of identity is an orthogonal one. The celebrated Naimark
dilation theorem (see Thm.~9.3.2 in \cite{dav}) says that for every
POVM $\map{M}{\Sigma_X}{\sB(\sH)}$ there exist a Hilbert space $\sK$,
a unitary $\map{U}{\sH}{\sK}$, a Hilbert space $\wdt{\sK}$
containing $\sK$ as a closed subspace, and a PVM
$\map{E}{\Sigma_X}{\sB(\wdt{\sK})}$, such that, for any $\Delta \in
\Sigma_X$,  $M(\Delta) = U^*PE(\Delta)PU$, where $P$ is the orthogonal
projection from $\wdt{\sK}$ onto $\sK$. Furthermore, we can
define the partial isometry $\map{V}{\sH}{\wdt{\sK}}$ (with the final
projection $P$) by $V = PU$, so that
$M(\Delta) = V^*E(\Delta)V$ \cite{note2}.

With these lengthy preliminaries out of the way, we can
proceed to state and prove our result.

\begin{theorem}\label{th:ordops} Consider quantum operations $T_i \in
  \cp(\sH_1,\sH_2)$, $i = 1,\ldots,n$, that satisfy $T_1 \le T_2 \le
  \ldots \le T_n$.  Then there exist a Hilbert space $\sH$, an
  isometry $\map{V}{\sH_2}{\sH_1 \tp \sH}$, and orthogonal projections
  $\Pi_i \in \sB(\sH)$ such that
\begin{enumerate}
\item $T_i(A) = V^*(A \tp \Pi_i)V$, $1 \le i \le n$
\item $\Pi_1 \le \Pi_2 \le \ldots \le \Pi_n$.
\end{enumerate}
Conversely, if items 1 and 2 above hold for quantum operations $T_i \in
\cp(\sH_1,\sH_2)$ with some $\sH$, $V$, and $\set{\Pi_i}$, then $T_1
\le T_2 \le \ldots \le T_n$.
\end{theorem}

\begin{proof} Suppose that $\set{T_i}$ satisfy the hypothesis of the
  theorem. Without loss of generality we may take $T_n$ to be a
  channel, for if not, then we can append to $\set{T_i}^n_{i=1}$ the channel $T_{n+1}(A) = M^*AM + T_n(A)$, where
  $\map{M}{\sH_2}{\sH_1}$ is an operator defined, up to a unitary,
  through $M^*M = \idty - T_n(\idty)$, so that the resulting collection
  $\set{T_i}^{n+1}_{i=1}$ still satisfies $T_1 \le T_2 \le \ldots \le T_{n+1}$.

Define quantum operations $S_i$, $i = 1,\ldots,n$, by
  $S_1 = T_1$ and $S_i = T_i - T_{i-1}$, $1 < i \le n$. Then $T_k =
  \sum^k_{i=1}S_i$, $1 \le k \le n$. If $T_n(A) = W^*(A \tp
  \idty_\sE)W$ is the canonical Stinespring dilation of $T_n$,
  Theorem \ref{th:rnsum} states that there exist positive operators
  $F_i \in \sB(\sE)$ such that $S_i(A)= W^*(A\tp F_i)W$, and $\sum_i
  F_i = \idty_\sE$. By the Naimark
  dilation theorem there exist a Hilbert space $\sH$, an
  isometry $\map{\wdt{V}}{\sE}{\sH}$, and a PVM $\set{E_i}^n_{i=1}$
  $,E_i \in \sB(\sH)$, such that $F_i = \wdt{V}^*E_i\wdt{V}$, $1 \le i
  \le n$. Thus we can write $S_i(A) = V^*(A \tp E_i)V$, where the isometry
  $\map{V}{\sH_2}{\sH_1 \tp \sH}$ is defined by $V =
  (\idty_{\sH_1} \tp \wdt{V})W$.

For each $k$, $1 \le k \le n$, let $\Pi_k = \sum^k_{i=1}E_i$. Since
$\set{E_i}$ is an orthogonal resolution of identity, each $\Pi_k$ is
an orthogonal projection, and $\Pi_k \le \Pi_l$ for $k \le
l$ by construction. Furthermore,
$$
T_k(A) = \sum^k_{i=1}S_i(A) = \sum^k_{i=1}V^*(A \tp E_i)V = V^*(A \tp
\Pi_k)V \qquad 1 \le k \le n,
$$
and the forward direction is proved.  The proof of the reverse
direction is straightforward.
\end{proof}

It is pertinent to remark that there are situations when the
correspondence between POVM's with values in a suitable Hilbert space
and decompositions of a given quantum channel $T$ into completely
positive summands is not merely a nice mathematical device, but in
fact acquires direct physical significance. For instance, Gregoratti and
Werner \cite{gw} have exploited this correspondence in a scheme for
recovery of classical and quantum information from noise by making a
generalized quantum measurement (described by a POVM \cite{hel}) on
the ``environment'' Hilbert space of a noisy quantum channel [the
  Hilbert space $\sE$ in the ``ancilla'' form (\ref{eq:ancilla})].

\section{Characterization of quantum operations by positive operators}
\label{sec:cqopo}

The correspondence between linear maps from a matrix algebra $\sM_m$
into a matrix algebra $\sM_n$ and linear functionals on $\sM_n \tp
\sM_m$ (or, by the Riesz lemma, linear opreators on $\C_n \tp \C_m$)
has been treated extensively in a variety of forms in the mathematical
literature (see, e.g., \cite{ls,cho,jam,pil,ph} for a sampling of
results related to positive and completely positive maps). More
recently, this correspondence has been exploited fruitfully in some
quantum information-theoretic contexts, such as optimal cloning maps
\cite{dp}, optimal teleportation protocols \cite{hhh}, separability
criteria for entangled states \cite{mm}, or entanglement generation
\cite{cdkl,zan}. In this section we will show that the one-to-one
correspondence between positive operators on $\C_n \tp \C_m$ and CP
maps $\map{T}{\sM_m}{\sM_n}$ (known as the ``Jamiolkowski
isomorphism'' in the quantum information community) can be derived
using the Radon-Nikodym machinery.  We also comment on how this can be
accomplished in the infinite-dimensional case with unbounded operators.

\subsection{The Jamiolkowski isomorphism}
\label{ssec:jam}

In this section we consider quantum operations
$\map{T}{\sB(\sH)}{\sB(\sK)}$ in the case of $\dim \sH = m < \infty$ and $\dim
\sK = n < \infty$. Let $\set{e_i}^m_{i=1}$ and
$\set{f_\mu}^n_{\mu=1}$ be fixed orthonormal bases of $\sH$ and $\sK$.  (We
will use Latin indices for the ``input'' Hilbert space, and Greek ones
for the ``output'' Hilbert space.)  Let $\tau$ be the tracial state on
$\sM_m$, $\tau(A) = m^{-1}\tr A$, and consider the channel $\Phi(A)
\defeq \tau(A)\idty_\sK$. It is convenient to write $\Phi$ in the Kraus form
$$
\Phi(A) \defeq \sum^m_{i=1}\sum^n_{\mu=1} V^*_{i\mu}AV_{i\mu},
$$
where $V_{i\mu} = \frac{1}{\sqrt{m}}\ketbra{e_i}{f_\mu}$. Note that these $mn$ Kraus operators are linearly independent, which
agrees with the minimality requirement. Setting $\sE
= \sK \tp \sH$, we obtain the canonical Stinespring dilation $\Phi(A) =
V^*_\Phi(A \tp \idty_\sE)V_\Phi$, where
$$
V_\Phi \psi = \sum^m_{i=1}\sum^n_{\mu=1}V_{i\mu}\psi \tp f_\mu \tp
e_i.
$$
Whenever we need to specify the
dimensions $m$ and $n$ explicitly, we will write
$\Phi_{m,n}$ instead of $\Phi$, $V_{m,n}$ instead of $V_\Phi$, etc. 

We must emphasize again that the main result of this section, stated
as Theorem \ref{th:jam} below, is not new. Indeed, it has appeared in
numerous papers on quantum information theory
\cite{dp,hhh,mm,cdkl,zan}. Our contribution here is to present a new
proof of this result that clearly exhibits the Jamiolkowski
isomorphism in the Radon-Nikodym framework.

\begin{theorem}\label{th:jam} In the notation described above, any CP map
  $\map{T}{\sB(\sH)}{\sB(\sK)}$ is completely $m^2$-dominated by
  $\Phi$. There exists a unique operator $F_T \in \sB(\sE)$ with $0
  \le F_T \le m^2 \idty_\sE$, such that $\dd_\Phi T = \idty_\sH \tp
  F_T$, i.e., $T(A) = V^*_\Phi(A \tp F_T)V_\Phi$. The action of $T$ on
  any $A \in \sB(\sH)$ can also be expressed in terms of $F_T$ only, namely as
\begin{equation}
T(A) = \frac{1}{m}\tr_\sH[(\idty_\sK \tp \tran{A})F_T],
\label{eq:jam}
\end{equation}
where $\tran{A}$ denotes the matrix transpose of $A$ in the basis
$\set{e_i}$. Furthermore, $T$ is a quantum operation if and only if
$\tr_\sH F_T \le m \idty_\sK$. 
\end{theorem}

\begin{proof} Define $\Psi = \frac{1}{\sqrt{m}}\sum^m_{i=1} e_i \tp e_i$, and
  let $H_T = T \tp \id (\ketbra{\Psi}{\Psi})$. The matrix elements
  of $H_T$ are given explicitly by
$$
\braket{f_\mu \tp e_i}{H_T(f_\nu \tp e_j)} = \frac{1}{m}
\braket{f_\mu}{T(\ketbra{e_i}{e_j})f_\nu}.
$$
For all $A \in \sB(\sH)$ and $\psi \in \sK$ we have
\begin{eqnarray*}
V^*_\Phi(A \tp H_T)V_\Phi \psi &=& \frac{1}{m}
\sum^m_{i,j=1}\sum^n_{\mu,\nu=1} \braket{e_i}{Ae_j} \braket{f_\mu \tp
  e_i}{H_T(f_\nu \tp e_j)}\ket{f_\mu}\braket{f_\nu}{\psi}\\
&=& \frac{1}{m^2}\sum^n_{\mu,\nu=1}\braket{e_i}{Ae_j}
\braket{f_\mu}{T(\ketbra{e_i}{e_j})f_\nu}\ket{f_\mu}\braket{f_\nu}{\psi}
\\
&\equiv& \frac{1}{m^2} T(A)\psi,
\end{eqnarray*}
so that $T \le m^2\Phi$ and $\idty_\sH \tp m^2 H_T = \dd_\Phi T$ by
Corollary \ref{cor:rn}. Let
$F_T = m^2 H_T$. From the uniqueness of
the Radon-Nikodym derivative $\dd_\Phi T$ it follows that $F_T$
determines $T$ uniquely.

To prove Eq.~(\ref{eq:jam}), we need the following useful identity.

\begin{lemma} \label{lm:tranid} For all $A \in \sB(\sH)$ and $B \in
    \sB(\sE)$, we have
$$
V^*_\Phi(A \tp B)V_\Phi = \frac{1}{m}
    \tr_\sH[(\idty_\sK \tp \tran{A})B].
$$
\end{lemma}

\begin{proof} Proceed by direct computation; for an arbitrary $\psi
  \in \sK$, we have
\begin{eqnarray*}
\tr_\sH[(\idty_\sK \tp \tran{A})B]\psi &=& \left(\tr_\sH \sum^m_{i,j,k=1}\sum^n_{\mu,\nu=1}\braket{e_j}{Ae_i}
\braket{f_\mu\tp e_j}{B(f_\nu \tp e_k)} \ketbra{f_\mu}{f_\nu} \tp
\ketbra{e_i}{e_k}\right)\psi \\
&=& \sum^m_{i,j=1}\sum^n_{\mu,\nu=1}\braket{e_j}{Ae_i}\braket{f_\mu
  \tp e_j}{B(f_\nu \tp e_i)}\ket{f_\mu}\braket{f_\nu}{\psi} \\
&\equiv & mV^*_\Phi(A \tp B)V_\Phi \psi,
\end{eqnarray*}
and the lemma is proved.
\end{proof}

\noindent{This establishes Eq.~(\ref{eq:jam}). Finally, if $T$ is a quantum
operation, then $T(\idty_\sH) \le \idty_\sK$.  From Lemma
\ref{lm:tranid} it follows that $T(\idty_\sH) = \frac{1}{m} \tr_\sH F_T$,
that is, $\tr_\sH F_T \le m\idty$. Conversely, if
$T(\idty_\sH) = V^*_\Phi(\idty_\sH \tp F_T)V_\Phi \le \idty_\sK$, we
have $\tr_\sH F_T \le m \idty_\sK$ by Lemma \ref{lm:tranid}. The
theorem is proved.}
\end{proof}

Let $T(A) = V^*(A \tp \idty_\sF)V$ be the canonical Stinespring
dilation of $T$. Then it is easily shown that $\dim \sH \cdot
\dim \sF = \rank \dd_\Phi T$, that is $\dim \sF = \rank F_T$. Indeed,
Theorems \ref{th:rnkraus2} and \ref{th:jam} together imply that for any CP map
$\map{T}{\sM_m}{\sM_n}$ there exist operators $\set{K^T_i}^N_{i=1}$
from $\sM_n$ into $\sM_m$,
such that
$\Phi_{m,n}(A) = \sum^N_{i=1} (K^T_i)^*AK^T_i$ and $T(A) = \sum^N_{i=1}
\lambda_i (K^T_i)^*AK^T_i$, where $\set{\lambda_i}$ are the (nonnegative)
eigenvalues of $F_T$. The Kraus operators $\set{K^T_i}^N_{i=1}$ are
linearly independent, and are determined
by the isometry $V_\Phi$ and the eigenvectors $\set{\xi_i}^N_{i=1}$ of
$F_T$ through $V_\Phi \psi = \sum^N_{i=1} V^T_i\psi \tp \xi_i$. Therefore
$N \equiv mn$. The number of nonzero terms in the corresponding Kraus
decomposition of $T$ is equal to $\rank F_T$, so that $\dim \sF = \rank F_T$.

Lastly we would like to show how the Radon-Nikodym derivative
$\dd_\Phi T$ transforms under composition of CP maps. Consider two CP
maps $\map{T_1}{\sM_m}{\sM_n}$ and
$\map{T_2}{\sM_n}{\sM_d}$. According to Theorem \ref{th:jam}
we can write
$$
T_1 (A) = V^*_{m,n}(A \tp F_1)V_{m,n}, \qquad T_2(B) = V^*_{n,d}(B \tp
F_2)V_{n,d}
$$
for uniquely determined positive operators $F_1 \in \sM_n \tp \sM_m$
and $F_2$ on $\sM_d \tp \sM_n$. For any $A \in \sM_m$, we have
\begin{eqnarray*}
T_2\circ T_1(A) &=& V^*_{n,d}(T_1(A) \tp F_2)V_{n,d} \\
&=& V^*_{n,d} \left( V^*_{m,n}(A \tp F_1)V_{m,n} \tp F_2\right)V_{n,d}
\\
&=& V^*_{n,d}(V^*_{m,n} \tp \idty_{d\times n})(A \tp F_1 \tp F_2)(V_{m,n} \tp
\idty_{d\times n})V_{n,d},
\end{eqnarray*}
where $\idty_{d\times n}$ denotes the identity operator on the dilation space
$\C^d \tp C^n$ of $T_2$. Let $\set{e_i}^m_{i=1}$, $\set{f_\mu}^n_{\mu
  = 1}$, and $\set{\phi_x}^d_{x=1}$ be orthonormal bases of $\C^m$,
  $\C^n$, and $\C^d$ respectively.  Then for any $A \in \sM_m$ and any
$\psi \in \C^d$ we have
\begin{eqnarray*}
T_2\circ T_1(A)\psi &=& \frac{1}{mn} \sum^m_{i,j=1} \sum^n_{\mu,\nu=1}
\sum^d_{x,y=1}\braket{e_i}{Ae_j}
\ket{\phi_x}\braket{\phi_y}{\psi} \\
&& \qquad \times \braket{f_\mu \tp e_i}{F_1(f_\nu \tp e_j)}
\braket{\phi_x \tp f_\mu}{F_2(\phi_y \tp f_\nu)} \\
&=& \frac{1}{m} \sum^m_{i,j=1} \sum^d_{x,y=1} \left(\frac{1}{n}
\sum^n_{\mu,\nu=1} \braket{\phi_x \tp f_\mu}{F_2(\phi_y \tp f_\nu)} \braket{f_\mu \tp e_i}{F_1(f_\nu \tp e_j)}
\right ) \\
&& \qquad \times \braket{e_i}{Ae_j}
\ket{\phi_x}\braket{\phi_y}{\psi}.
\end{eqnarray*}
Let $\Omega = \frac{1}{\sqrt{n}} \sum^n_{\mu=1}f_\mu \tp f_\mu$. Define an
operator $F_{21}$ on $\C^d \tp C^m$ by
$$
\braket{\phi_x \tp e_i}{F_{21}(\phi_y \tp e_j)} = \braket{\phi_x \tp
  \Omega \tp e_i}{(F_2 \tp F_1)(\phi_y \tp \Omega \tp e_j)}.
$$
Then it is evident from the calculations above that we can write
$T_2\circ T_1(A) = V^*_{m,d}(A \tp F_{21})V_{m,d}$. By the uniqueness of the
Radon-Nikodym derivative, $\idty_m \tp F_{21} = \dd_{\Phi_{m,d}}(T_2
\circ T_1)$. Defining the conditional expectation $M_\Omega$ from
$\sM_d \tp \sM^{\tp 2}_n \tp \sM_m$ onto $\sM_d \tp \sM_m$ by
$$
M_\Omega(A \tp B \tp C) = \braket{\Omega}{B\Omega}(A \tp C)
\qquad \all A \in \sM_d, B \in \sM^{\tp 2}_n, C \in
\sM_m,
$$
we can write more succinctly $F_{21} = M_\Omega(F_2 \tp F_1)$.

\subsection{Generalization to arbitrary faithful states}
\label{ssec:faith}

The construction described in
Sect.~\ref{ssec:jam} also goes through
if, instead of the tracial state $\tau$, we take an arbitrary {\em
  faithful} state $\omega$. As is well-known, for any such state there
exist an orthonormal basis $\set{e_i}^m_{i=1}$ and a probability distribution
$\set{p_i}^m_{i=1}$ with $p_i > 0$, such that $\omega(A) =
\sum^m_{i=1}p_i \braket{e_i}{Ae_i}$ for all $A \in
\sB(\sH)$. Furthermore, $\omega(A) = \braket{\Omega}{(A \tp
  \idty)\Omega}$, where $\Omega = \sum^m_{i=1}\sqrt{p_i} e_i \tp
e_i$. (This is, of course, the canonical Stinespring dilation
of the CP map $\omega$ by means of the GNS construction.) Let
$D_\omega \in \sB(\sH)$ denote the density operator corresponding to
$\omega$, i.e., $\omega(A) = \tr(D_\omega A)$. Owing to the
faithfulness of $\omega$, $D_\omega$ is invertible.

Fix an orthonormal basis $\set{f_\mu}^n_{\mu=1}$ of $\sK$, and define
the channel $\map{\Phi_\omega}{\sB(\sH)}{\sB(\sK)}$ through
$\Phi_\omega(A) = \omega(A)\idty_\sK$. The Kraus form of $\Phi_\omega$
is given by $\Phi_\omega(A) = \sum^m_{i=1}\sum^n_{\mu=1}
V^*_{i\mu}AV_{i\mu}$, where $V_{i\mu} = \sqrt{p_i}
\ketbra{e_i}{f_\mu}$, and the canonical Stinespring dilation by
$\Phi_\omega(A) = V^*_\omega (A \tp \idty_\sE)V_\omega$, where again
$\sE \simeq \sK \tp \sH$ and $V_\omega \psi =
\sum^m_{i=1}\sum^n_{\mu=1}V_{i\mu}\psi \tp f_\mu \tp e_i$. 

Consider the positive operator $F_{T,\omega} =
  T\tp\id\left((D^{-1}_\omega \tp
  \idty)\ketbra{\Omega}{\Omega}(D^{-1}_\omega \tp \idty)\right)$,
  whose matrix elements are given by $\braket{f_\mu \tp
  e_i}{F_{T,\omega}(f_\nu \tp e_j)} = \frac{1}{\sqrt{p_ip_j}}
  \braket{f_\mu}{T(\ketbra{e_i}{e_j})f_\nu}$. For all $A \in \sB(\sH)$
  and $\psi \in \sK$ we then have
\begin{eqnarray*}
V^*_\omega (A \tp F_{T,\omega})V_\omega \psi &=&
\sum^m_{i,j=1}\sum^n_{\mu,\nu=1} \sqrt{p_i p_j}
\braket{e_i}{Ae_j} \braket{f_\mu
  \tp e_i}{F_{T,w}(f_\nu \tp e_j)} \ket{f_\mu}\braket{f_\nu}{\psi}\\
&=& \sum^m_{i,j=1}\sum^n_{\mu,\nu=1}\braket{e_i}{Ae_j}
\braket{f_\mu}{T(\ketbra{e_i}{e_j})f_\nu}\ket{f_\mu}\braket{f_\nu}{\psi}
\\
&\equiv& T(A)\psi,
\end{eqnarray*}
so that $T \le \norm{F_{T,\omega}}\Phi_\omega$, with $\norm{F_{T,\omega}}
\le \norm{D^{-1}_\omega}^2 \cbnorm{T}$.

Consequently, for any faithful state $\omega$ on $\sB(\sH)$ and any CP
map $\map{T}{\sB(\sH)}{\sB(\sK)}$ there exists a positive constant
$c$ such that $T$ is completely $c$-dominated by $\Phi_\omega$; thus $T$
is uniquely determined by the Radon-Nikodym derivative
$\dd_{\Phi_\omega}T$. Note that in the special case of $\omega$ being
the tracial state on $\sM_m$ we simply recover the results of the
preceding section.

\subsection{Generalization to infinite dimensions}
\label{ssec:infdim}

In the form stated above, both the Jamiolkowski isomorphism and its
generalization to arbitrary faithful states are valid only for CP maps between
finite-dimensional algebras.  However, in many problems of quantum
information theory it is necessary to consider CP maps between
algebras of operators on infinite-dimensional Hilbert spaces.

Consider a normal CP map $\map{T}{\sB(\sH)}{\sB(\sK)}$, where $\sH$
and $\sK$ are separable Hilbert spaces. Fix a normal faithful state
$\omega$ on $\sB(\sH)$; then there exist a complete orthonormal basis
$\set{e_i}$ of $\sH$ and a probability distribution $\set{p_i}$,
$p_i > 0$, such that, for any $A \in \sB(\sH)$, $\omega(A) =
\braket{\Omega}{(A \tp \idty)\Omega}$ with $\Omega = \sum_i
\sqrt{p_i}e_i \tp e_i$. Let
$D_\omega$ denote the density operator corresponding to
$\omega$. Because $\sH$ is infinite-dimensional, the inverse of
$D_\omega$ is an unbounded operator defined on a dense domain, namely
the linear span of
$\set{e_i}$. Therefore the approach taken in the preceding section will
not work; instead, we will characterize $T$ through the Radon-Nikodym
derivative of another CP map $T_\omega$ (dependent on both $T$ and
$\omega$) with respect to the channel $\Phi_\omega = \omega(A)\idty_\sK$.

Choosing a complete orthonormal basis $\set{f_\mu}$ of $\sK$, we can write
$\Phi_\omega$ in the Kraus form $\Phi_\omega(A) = \sum_{i,\mu}
V^*_{i\mu}AV_{i\mu}$, $V_{i\mu} = \sqrt{p_i}
\ketbra{e_i}{f_\mu}$, where the series converges in the strong operator topology. We also have the Stinespring dilation via $\Phi_\omega(A) = V^*_\omega (A \tp
\idty_\sE)V_\omega$, where $\sE \simeq \sK \tp \sH$ and
$V_\omega \psi = \sum_{i,\mu}V_{i\mu}\psi \tp f_\mu \tp
e_i$. To see that this Stinespring dilation is canonical, let $A =
\frac{1}{\sqrt{p_k}}\ketbra{e_j}{e_k}$ and $\psi = f_\nu$. Thus
$$
(A \tp \idty_\sE)V_\omega\psi = e_j \tp f_\nu \tp e_k,
$$
which shows that the set $\setcond{(A \tp \idty_\sE)V_\omega \psi}{A \in \sB(\sH),\psi
  \in \sK}$ is total in $\sH \tp \sE$.

Let $F_{T,\omega} = T\tp\id(\ketbra{\Omega}{\Omega})$; the matrix
elements are
$$
\braket{f_\mu \tp e_i}{F_{T,\omega}(f_\nu \tp
  e_j)} = \sqrt{p_ip_j}\braket{f_\mu}{T(\ketbra{e_i}{e_j})f_\nu}.
$$
Then for all $A
\in \sB(\sH)$ and $\psi \in \sK$ we can write
\begin{eqnarray*}
V^*_\omega(A \tp F_{T,\omega})V_\omega \psi &=&
\sum_{i,\mu}\sum_{j,\nu} \sqrt{p_i p_j}\braket{f_\mu \tp
  e_i}{F_{T,\omega}(f_\nu \tp e_j)}
\braket{e_i}{Ae_j}\ket{f_\mu}\braket{f_\nu}{\psi} \\
&=& \sum_{i,\mu}\sum_{j,\nu} p_i p_j
\braket{e_i}{Ae_j}\braket{f_\mu}{T(\ketbra{e_i}{e_j})f_\nu}\ket{f_\mu}\braket{f_\nu}{\psi}
\\
&\equiv& T(D_\omega A D_\omega)\psi.
\end{eqnarray*}
We will write $T_\omega(A)$ for
$T(D_\omega A D_\omega)$. From the Radon-Nikodym theorem it follows
that $T_\omega$ is completely dominated by $\Phi_\omega$, and that
$\dd_{\Phi_\omega}T_\omega = \idty_\sH \tp F_{T,\omega}$. We can
determine the action of $T$ on the ``matrix units'' $\ketbra{e_i}{e_j}$ via
$T(\ketbra{e_i}{e_j}) = (p_i p_j)^{-1} T_\omega(\ketbra{e_i}{e_j})$.

\section{Norm estimates for differences of quantum operations}
\label{sec:norms}

In this section we will demonstrate the use of the Radon-Nikodym
theorem for CP maps in deriving several useful estimates for CB norms of
differences of quantum channels.

Consider two CP maps $\map{T_1,T_2}{\sB(\sH)}{\sB(\sK)}$. Suppose that
  there exists a CP map $\map{T}{\sB(\sH)}{\sB(\sK)}$, such that $T_i
  \le T$, $i=1,2$, and let $T(A) = V^*(A \tp \idty_\sE)V$ be the
  canonical Stinespring dilation of $T$. By the
  Radon-Nikodym theorem, there exist positive contractions $F_1, F_2
  \in \sB(\sE)$ such that $T_i(A) = V^*(A \tp F_i)V$, $i=1,2$. Then
$$
(T_1 - T_2)(A) = T_1(A) - T_2(A) = V^*\left(A \tp (F_1 - F_2)\right)V,
$$
and the Haagerup-Paulsen-Wittstock theorem immediately implies that
$$
\cbnorm{T_1-T_2} \le \norm{V}\norm{(F_1 - F_2)V} \le \norm{V}^2
\norm{F_1 - F_2}.
$$
If $T$ is a quantum channel, $V$ is an isometry, so that
$\norm{V}=1$. Therefore we get
\begin{equation}
\cbnorm{T_1 - T_2} \le \norm{F_1 - F_2}.
\label{eq:rncbest1}
\end{equation}
In particular, if $S \le T$, then $\cbnorm{S-T} \le \norm{\idty-F}$,
where $\idty \tp F$ is the Radon-Nikodym derivative $\dd_T S$.

Given two CP maps $\map{T_1,T_2}{\sB(\sH)}{\sB(\sK)}$ with
(not necessarily minimal) Stinespring dilations $T_i(A) = V^*_i(A \tp \idty_\sE)V_i$, $i=1,2$, on the
common dilation space $\sE$, the norm $\cbnorm{T_1-T_2}$ can be
bounded from above in terms of $V_1$ and $V_2$. Indeed, denoting by
$\pi$ the $*$-homomorphism $\mapi{\sB(\sH) \ni A}{A \tp \idty_\sE}$,
we can use the Haagerup-Paulsen-Wittstock theorem to obtain
\begin{eqnarray}
\cbnorm{T_1 - T_2} &=& \cbnorm{V^*_1 \circ \pi \circ V_1 - V^*_2 \circ \pi
  \circ V_2} \nonumber \\
&\le& \cbnorm{V^*_1 \circ \pi \circ V_1 - V^*_1 \circ \pi \circ V_2} +
  \cbnorm{V^*_1 \circ \pi \circ V_2 - V^*_2 \circ \pi \circ V_2}
  \nonumber \\
&\le & (\norm{V_1}+\norm{V_2})\norm{V_1-V_2}.\label{eq:cbnormup}
\end{eqnarray}
If $T_1$ and $T_2$ are channels, then $V_1$ and $V_2$ are
isometries. Consequently, $\norm{V_1} = \norm{V_2} = 1$, and the bound
(\ref{eq:cbnormup}) becomes $\cbnorm{T_1 - T_2} \le 2\norm{V_1 -
  V_2}$. As the lemma below shows, when the Hilbert spaces $\sH$ and $\sK$ are
finite-dimensional, one can find a
common dilation space $\sE$ and maps $\map{V_1,V_2}{\sK}{\sH \tp
  \sE}$, such that $\cbnorm{T_1-T_2}$ can be bounded from below.

\begin{lemma}\label{lm:cbnormest} For any two CP maps
  $\map{T_1,T_2}{\sB(\sH)}{\sB(\sK)}$ there exist a Hilbert space
  $\sE$ and operators $\map{V_1,V_2}{\sK}{\sH \tp \sE}$ such
  that $T_i(A) = V^*_i (A \tp \idty_\sE)V_i$, $i=1,2$, and
\begin{equation}
\norm{V_1 - V_2} \le \dim \sH \sqrt{\cbnorm{T_1-T_2}}.
\label{eq:cbnormest}
\end{equation}
\end{lemma}

\begin{proof} Using Theorem \ref{th:jam}, we can write $\sE = \sK
  \tp \sH$ and $V_i = \sqrt{\dd_\Phi T_i}V_\Phi = (\idty_\sH \tp
  \sqrt{F_{T_i}})V_\Phi$. Then $T_i(A) = V^*_i(A \tp
  \idty_\sE)V_i$. Next we prove the estimate (\ref{eq:cbnormest}). We
  have
\begin{equation}
\norm{V_1 - V_2} \le \norm{\idty_\sH \tp \sqrt{F_{T_1}} - \idty_\sH
  \tp \sqrt{F_{T_2}}} \norm{V_\Phi} = \norm{\sqrt{F_{T_1}} -
  \sqrt{F_{T_2}}} \le \sqrt{\norm{F_{T_1} - F_{T_2}}}.
\label{eq:step1}
\end{equation}
The last inequality in (\ref{eq:step1}) holds because: (1) 
$\mapi{x}{\sqrt{x}}$ is an operator monotone function on $[0,\infty)$,
  i.e., $\sqrt{A} \le \sqrt{B}$ for all operators $A,B$ satisfying $0
  \le A \le B$ (Prop.~V.1.8 in \cite{bha}), (2) for any operator monotone function $f$
  with $f(0)=0$ and any pair of positive operators $A,B$ we have
  $\norm{f(A) - f(B)} \le f(\norm{A-B})$ (Thm.~X.1.1 in \cite{bha}), and (3) $\norm{X} = \norm{\sqrt{X}}^2$ for
  any $X\ge 0$ by
  the spectral mapping theorem. Now $F_{T_i} = (\dim \sH)^2 T_i \tp
  \id(\ketbra{\Psi}{\Psi})$, where $\Psi = (1/\sqrt{\dim \sH}) \sum_i
  e_i \tp e_i$ for some orthonormal basis $\set{e_i}$ in $\sH$. Thus,
  using the properties of the CB norm, we get
\begin{equation}
\norm{F_{T_1} - F_{T_2}} = (\dim \sH)^2 \norm{T_1 \tp
  \id(\ketbra{\Psi}{\Psi}) - T_2 \tp \id(\ketbra{\Psi}{\Psi})} \le
  (\dim \sH)^2 \cbnorm{T_1 - T_2}.
\label{eq:step2}
\end{equation}
Combining Eqs.~(\ref{eq:step1}) and (\ref{eq:step2}) yields
(\ref{eq:cbnormest}).
\end{proof}

Inequality (\ref{eq:cbnormest}) was also proved by Kitaev \cite{kit},
  but by quite different means. Here several
  warnings are in order.  In the article of Kitaev \cite{kit} the
  ``canonical representation'' of a CP map
  $\map{T}{\sB(\sH)}{\sB(\sK)}$ is defined as $T(A) = \tr_\sF WAW^*$
  with $\sF \simeq \sK \tp \sH$. This is not to be confused with the
  {\em canonical Stinespring dilation} of $T$, $T(A) = V^*(A \tp
  \idty_\sE)V$ [or its dual, $T_*(A) = \tr_\sE VAV^*$] which must satisfy
  the requirement that $\sH \tp \sE$ is (the closure of) the linear span of
  $\setcond{(A \tp \idty_\sE)V\psi}{A \in \sB(\sH), \psi \in
  \sK}$. Thus $\sE$ is, in general, a subspace of $\sF = \sK \tp
  \sH$. Furthermore, Kitaev's version of the estimate
  (\ref{eq:cbnormest}) has $\dim\sK$, and not $\dim\sH$, multiplying
  the CB norm on its right-hand side. This is due to the fact that,
  whereas we cast all CP maps in the Stinespring form $T(A) = W^*(A
  \tp \idty_\sF)W$, Kitaev prefers to work with the dual
  representation $T_*(A) = \tr_\sF WAW^*$. Since all (bounded)
  operators on a finite-dimensional Hilbert space are trace-class,
  $T_*$ trivially extends to a CP map from $\sB(\sK)$ into
  $\sB(\sH)$.

\section{Concluding remarks}
\label{sec:rems}

In this article we have shown that the Radon-Nikodym theorem for
completely positive maps \cite{arv,bs,par} is an extremely powerful
and versatile tool for problems involving characterization and
comparison of quantum operations. The upshot is that if $T(A) = V^*(A
\tp \idty_\sE)V$ is the canonical Stinespring dilation of a CP map
$T$, then the set of all CP maps $S$ for which $T-S$ is also CP (we
say that $S$ is {\em completely dominated} by $T$) is in a one-to-one
correspondence with the positive contractions $F$ on $\sE$, given
explicitly by $S(A) = V^*(A \tp F)V$. As we have demonstrated, this
correspondence brings many seemingly unrelated problems into a common
framework.

However, many important questions still remain unanswered. For
instance, it is not difficult to convert the above ``Stinespring form''
of the Radon-Nikodym theorem into an equivalent ``Kraus form''
(cf. Sect.~\ref{ssec:rnkra}). The Kraus decomposition of a CP
map $T$ involves at most countably many terms, and all maps $S$ completely
dominated by $T$ can be characterized in terms of positive-definite
kernels on the corresponding indexing set. However, it is not clear how to apply this theorem
directly to CP maps given in terms of a ``continual'' Kraus
decomposition (as in, e.g., the quantum operational model of Gaussian
displacement noise \cite{hall}). For example, if $U_g$ is a strongly continuous unitary
representation of a compact topological group $G$ on a Hilbert space
$\sH$, how do we describe all CP maps completely dominated by the channel
$$
T(A) = \int_G U^*_g A U_g d\mu(g),
$$
where $\mu$ is the (normalized) Haar measure on $G$, in terms of
$\set{U_g}$? A partial step in this direction has been taken by
Parthasarathy \cite{par}, who constructed a Stinespring dilation of
$T$ in terms of $\set{U_g}$ under the assumption that these operators
are linearly independent $\mu$-almost everywhere, i.e., 
$$
\int_G \varphi(g)U_g d\mu(g) = 0 \quad \Longleftrightarrow \quad
\varphi(g)=0 \,\, \mu.{\rm -a.e.}
$$
for any $\varphi \in \sL^1(G,\mu)$. However, a general solution is
still lacking. We hope to address this issue in a future publication.

\section*{Acknowledgments}

The author wishes to thank V.P. Belavkin for introducing him to the
subject of this paper, and G.M. D'Ariano for stimulating
discussions and a careful reading of the manuscript. This work was
supported by the U.S. Army Research Office through MURI grant
DAAD19-00-1-0177, and by the Defense Advanced Research Projects
Agency through QuIST grant F30602-01-2-0528.



\begin{thebibliography}{99}
\bibitem{note1}Strictly
  speaking, one should start with the \Schrodinger-picture map
  $T_*$, extend it by linearity to the trace class $\sT(\sH)$, and then
  {\it define} the Heisenberg-picture map $T$ by noting that, for a
  fixed $A \in \sB(\sH)$, $\tr[T_*(\rho)A]$ is a continuous linear
  functional on the trace class, and therefore has the form $\tr (\rho
  A')$ for some $A' \in \sB(\sH)$.  We then set $T(A) = A'$.  However,
  since we have assumed at the outset that $T$ is normal, we do not
  have to worry too much about this nuance.
\end{thebibliography}
\end{document}